\documentclass[12pt]{article}
\usepackage{graphicx}  % needed for figures  
\usepackage{dcolumn}   % needed for some tables   
\usepackage{bm}        % for math   
\usepackage{amssymb}   % for math  
\usepackage{amsmath}   
\usepackage{mathrsfs}  %for \mathscr       
\usepackage{graphicx,color} %for color     
    
\usepackage{abstract}
\usepackage[english]{babel}      
\usepackage{graphicx,wrapfig,lipsum}        
\usepackage{multimedia}    
\usepackage[mathscr]{eucal}   
\usepackage{bm}              
\usepackage{amssymb}        
\usepackage{amsmath}    
\usepackage{mathrsfs}       
\usepackage{sidecap} 
\usepackage{wrapfig}    
\usepackage{subfig} 
\usepackage{float}   
\begin{document}
%%%%%%%%%%%%%%%%%%%%%%%%%%%%%%%%%%%%%%%%%%%%%%%%%%%%%%%%%%%%%%%%%%%%%%%%%%%
\newpage
\title{Quantum effects due to a moving Dirichlet point}  
\author{C.~D.~Fosco$^{a}$, D. R. ~Junior$^{b}$, and
L.~E.~Oxman$^{b}$\\ 
{\normalsize\it $^a$Centro At\'omico Bariloche and Instituto Balseiro}\\ 
{\normalsize\it Comisi\'on Nacional de Energ\'\i a At\'omica}\\ 
{\normalsize\it R8402AGP Bariloche, Argentina.}\\
{\normalsize\it $^b$ Instituto de F\'{\i}sica}\\
{\normalsize\it Universidade Federal Fluminense}\\
{\normalsize\it Campus da Praia Vermelha}\\ 
{\normalsize\it Niter\'oi, 24210-340, RJ, Brazil.}} 
\date{}  
\maketitle 
\begin{abstract} 
	We study quantum effects induced by a point-like object that imposes Dirichlet boundary conditions along its world-line, on 
	a real scalar field $\varphi$ in 1, 2 and 3 spatial dimensions. The boundary conditions result from the strong coupling limit of a 
	term quadratic in the field and localized on the particle's trajectory. 
	We discuss the renormalization issues that appear and evaluate the effective action. Special attention is paid to the case 
	of 2 spatial dimensions where the coupling constant is adimensional.
\end{abstract}
%%%%%%%%%%%%%%%%%%%%%%%%%%%%%%%%%%%%%%%%%%%%%%%%%%%%%%%%%%%%%%%%%%%%%%%%%%
%%%%%%%%%%%%%%%%%%%%%%%%%%%%%%%%%%%%%%%%%%%%%%%%%%%%%%%%%%%%%%%%%%%%%%%%%% 
%%%%%%%%%%%%%%%%%%%%%%%%%%% Introduction %%%%%%%%%%%%%%%%%%%%%%%%%%%%%%%%%
%%%%%%%%%%%%%%%%%%%%%%%%%%%%%%%%%%%%%%%%%%%%%%%%%%%%%%%%%%%%%%%%%%%%%%%%%%
%%%%%%%%%%%%%%%%%%%%%%%%%%%%%%%%%%%%%%%%%%%%%%%%%%%%%%%%%%%%%%%%%%%%%%%%%%
\section{Introduction}\label{sec:intro}
Quantum Field Theory predicts that an open system is capable of evolving,
from the vacuum, to a state characterized by a non-vanishing number of
(real) quanta~\cite{moore}.  That is indeed the case, among other related
phenomena, of the Dynamical Casimir Effect (DCE), one of the most studied
manifestations of  quantum dissipation~\cite{reviews}.  
The DCE consists of the emission of {\em real\/} quanta when a field is subjected
to time-dependent boundary conditions, an example being the
presence of one or more moving mirrors, namely, of objects imposing
non-trivial boundary conditions on the field. In the usual understanding of
the term, a boundary condition acts on a region having co-dimension one,
i.e., which is determined by a single equation.  It is worth noting that,
in the context of the DCE for a real scalar field, which we consider here,
different kinds of boundary conditions, besides the `perfect' ones
(Dirichlet and Neumann),  have also been studied. 
Those `imperfect' conditions describe mirrors which have more realistic
responses to the action of the field's modes.  Among that kind of
condition, a relatively simple one amounts to Dirichlet-like boundary
conditions: they result from the addition to the action of a term
localized on the space-time region which is swept by the mirror during the
course of time. When the strength of that term tends to infinity, one gets
Dirichlet conditions on the region on which the term is localized.
It is our concern in this paper to study the DCE, for the case of 
a real scalar field $\varphi$ in $d+1$ dimensions ($d=1,2,3$), coupled to
point-like objects which implement precisely that kind of Dirichlet-like
boundary conditions. In other words, we shall add to the scalar field
Lagrangian a term proportional to a $\delta$-function of
the (time-dependent) position of the particle, and to the square of
$\varphi$.  The strength of the term is determined by a coupling
constant which, by taking the appropriate limit,  will be used to impose
Dirichlet boundary conditions.

We shall follow our previous work for scalar and spinorial vacuum
fields~\cite{prd07,prd11} in which we used  the particularly convenient
functional approach proposed by Golestanian and Kardar~\cite{GK}. The
approach is based on the use  of auxiliary fields to deal with the role of
the mirrors, on the calculation of the functional integral for the in-out
effective action.
An important feature of the systems that we consider here is the following: 
except for $d=1$  a curve, like the particle's world-line, has codimension
bigger than $1$; this fact results in qualitatively different UV properties
during the calculation of the effective action. Indeed, the UV problems
which will arise here are rather similar to the ones corresponding to Dirac
$\delta$-potentials in $2$ and $3$ dimensions, a system which has been
extensively studied by following many different approaches and frameworks
(see, for example,
\cite{Berezin:1960df,Albeverio:1976xf,Jackiw:1991je,Camblong:2000qn}).
Note that the classical, static Casimir effect for small objects is one of
the problems considered in~\cite{small}, by using a multipole expansion.
What we have in mind here is the evaluation of the dynamical and  quantum
version of that kind of object.

In this paper, we shall first review the $d=1$ case, as a previous step to
dealing with $d=2$, and $d=3$.  The main distinction between $d=2$ or $d=3$
and $d=1$ are, as we shall see, due to the different UV properties induced by the 
coupling between the particle and the field. Indeed, the usual
renormalization which is required to make sense of a $\delta$-like
potential in Quantum Mechanics in two and three spatial dimensions, also
manifests itself here; moreover, the resulting divergences can be cured by
applying a similar procedure.  

The $d=2$ case will also be relevant for future developments regarding the
quantum properties of center vortices, and support the construction of
phenomenological ensembles for these magnetic defects. They are topological
variables that are believed to capture the infrared behavior in Yang-Mills
theories.  For a given realization, the calculation of the effective action
involves regularity field conditions on world-lines and world-surfaces in
three and four Euclidean dimensions, respectively, which are  problems of
codimension 2.  The effective action for a single center vortex without
curvature was analyzed in Refs.~\cite{niel1}-\cite{oxman}. This may involve
singular spectral problems with different gyromagnetic ratios and
regularity conditions on the YM off-diagonal sector, which depend on how
the fluctuations are parametrized.

The structure of this paper is as follows: in Sect.~\ref{sec:thesys} we
introduce the kind of system that we study, as well as some general
expressions for its effective action in the small departure limit, in the path integral framework.
Then, in Sect.~\ref{sec:pert} we evaluate the effective action for the massless field, by
considering a perturbative expansion in powers of the departure of the
worldline from the one of a static particle, assuming the mirror moves
non-relativistically, for $d=1$ and $d=3$.  
Because of its particularities, related to scale invariance of the coupling
between field and mirror, the $d=2$ case is considered separately in
Sec.~\ref{sec:par}. Finally, in Sec.~\ref{sec:conc} we present our conclusions. 

%%%%%%%%%%%%%%%%%%%%%%%%%%%%%%%%%%%%%%%%%%%%%%%%%%%%%%%%%%%%%%%%%%%%%
%%%%%%%%%%%%%%%%%%%%%%%%%%%%%%%%%%%%%%%%%%%%%%%%%%%%%%%%%%%%%%%%%%%%%
%%%%%%%%%%%%%%%%%%%%%%%%%% The system %%%%%%%%%%%%%%%%%%%%%%%%%%%%%%%
%%%%%%%%%%%%%%%%%%%%%%%%%%%%%%%%%%%%%%%%%%%%%%%%%%%%%%%%%%%%%%%%%%%%%
%%%%%%%%%%%%%%%%%%%%%%%%%%%%%%%%%%%%%%%%%%%%%%%%%%%%%%%%%%%%%%%%%%%%%
\section{The system}\label{sec:thesys}
 The system that we shall deal with throughout this paper consists of a
 real scalar field $\varphi$ in $d+1$ dimensions, with $d=1,\, 2,\, {\rm
 or}\, 3$, coupled to point-like objects which are meant to implement
 Dirichlet-like conditions.  

 For an object imposing Dirichlet conditions, the effective action will be
 denoted by $\Gamma({\mathcal C})$, since it is a functional of the world-line ${\mathcal C}$. In
 a functional integral approach,  and using Euclidean
 conventions~\footnote{A Wick rotation back to real time will be performed
 afterwards when dealing with the calculation of its imaginary part.}, 
\begin{equation}
	e^{- \Gamma({\mathcal C})} \;=\; 
\frac{{\mathcal Z}({\mathcal C})}{{\mathcal Z}_0} \;,
\end{equation}
where ${\mathcal Z}({\mathcal C})$ (${\mathcal Z}_0$) denotes the Euclidean
vacuum transition amplitude corresponding to the scalar field in the
presence (absence) of the particle.  

${\mathcal Z}({\mathcal C})$ and ${\mathcal Z}_0$ are given, explicitly, by
\begin{equation}\label{eq:defzc}
{\mathcal Z}({\mathcal C}) \;=\;  \int {\mathcal D}\varphi \,
\delta_{\mathcal C}(\phi)  \; e^{-{\mathcal S}_0(\varphi)} \;\;,\;\;\;
{\mathcal Z}_0 \;=\;  \int {\mathcal D}\varphi 
\; e^{-{\mathcal S}_0(\varphi)}\;,
\end{equation}
where ${\mathcal S}_0$ is the action which describes the free propagation
of the field, and a functional $\delta$-function has been introduced to
account for the Dirichlet conditions; namely, the vanishing of the field at
the position of the particle.  The former is given by: 
\begin{equation}\label{eq:defs}
{\mathcal S}_0(\varphi) \;=\; \frac{1}{2} \, \int_x \,\left(
\partial_\mu \varphi(x) \partial_\mu \varphi(x) +m^2\varphi^2(x)\right)\;,
\end{equation}
where we have introduced a shorthand notation for the integration, in this
case over all of the spacetime coordinates $x = (x_0, x_1, \ldots, x_d)$.
Namely, in the case above, \mbox{$\int_x \equiv \int d^{d+1}x$}.
Greek indices will be assumed to run over the values $0, 1, \ldots, d$, and
space-time is endowed with the Euclidean metric: $g_{\mu\nu} =
\delta_{\mu\nu}$. 

Regarding $\delta_{\mathcal C}(\phi)$, it should select, among the
configurations appearing in the functional integration measure, just the
$\varphi$-field configurations which satisfy Dirichlet boundary conditions
on ${\mathcal C}$.  As already advanced, those conditions will be reached
as the limit of a local term, namely: we add to the free action a term,
quadratic in $\varphi$ and localized on ${\mathcal C}$, with a strength
$\lambda$ which, when $\lambda \to \infty$, imposes Dirichlet boundary
conditions:  
\begin{equation}
\Gamma({\mathcal C}) \;=\; \lim_{\lambda \to \infty}
	\Gamma_\lambda({\mathcal C}) \;,\;\;\;
	e^{-\Gamma_\lambda({\mathcal C})} \;=\; {\mathcal
	Z}_\lambda({\mathcal C})/{\mathcal Z_0}
\end{equation}
where
\begin{equation}
{\mathcal Z}_\lambda({\mathcal C}) \;=\; \int {\mathcal D}\varphi
\; \exp\{-{\mathcal S}_0(\varphi)  -\frac{\lambda}{2}\int_\tau \,
\sqrt{g(\tau)} \, [\varphi (y(\tau))]^2\} 
\end{equation}
where we have assumed that $\tau \, \to \, y_\mu(\tau)$ ($\mu= 0,1,\ldots,
d$) is a parametrization of ${\mathcal C}$, and $g(\tau) \equiv
\dot{y}_\mu(\tau)  \dot{y}_\mu(\tau)$  (we have ignored, as customary in
the functional integral context, irrelevant factors which in this case are
independent of the curve and the field). The $\sqrt{g(\tau)}$ factor has
been introduced in order to have reparametrization invariance.

It is rather convenient to use an auxiliary field $\xi(\tau)$, in order to
have an alternative representation for the functional above, where
$\varphi$ may be integrated out in a simpler fashion. Indeed,
\begin{align}
{\mathcal Z}_\lambda({\mathcal C}) \;= & \int {\mathcal D}\varphi \, {\mathcal D}\xi  \; 
e^{-\frac{1}{2} \int_x \left(\partial_\mu\varphi \partial_\mu\varphi+m^2\varphi^2\right)  + i \int_x
J_{\mathcal C} \varphi - \frac{1}{2\lambda} \int_\tau\sqrt{g(\tau)}
[\xi(\tau)]^2} \;,
\end{align}
where
\begin{equation}
J_{\mathcal C}(x)\;\equiv\; \int_\tau\sqrt{g(\tau)} \xi(\tau)
\delta(x-y(\tau)) \;.
\end{equation}

Integrating out $\varphi$, we see that: 
\begin{equation}\label{eq:zc}
{\mathcal Z}_\lambda({\mathcal C}) \;=\;  {\mathcal Z}_0 \; \int\, {\mathcal D}\xi  \; 
e^{-\frac{1}{2} \int_{\tau, \tau'} \xi(\tau) \, {\mathcal K}(\tau,\tau') \xi(\tau')} \;.
\end{equation}
with a kernel ${\mathcal K}$ which may be rendered as follows:
\begin{align}
	{\mathcal K}(\tau,\tau') \;=\; \sqrt{g(\tau)}\, \Big[
	&\frac{\delta(s(\tau)-s(\tau'))}{\lambda}
	\nonumber\\
	+\, \langle y(\tau) |& (-\partial^2+m^2)^{-1} |  y(\tau') \rangle
	\Big] \, \sqrt{g(\tau')}\;, 
	\label{Ker}
\end{align}
where $s(\tau)$ denotes the Euclidean version of the proper time, namely,
the arc length, and we have used a bra-ket notation for the kernel of an
operator.

Then, integration of the auxiliary field yields
\begin{equation}\label{eq:zcdet}
{\mathcal Z}_\lambda({\mathcal C}) 
	\;=\; {\mathcal Z}_0 \; \big( \det {\mathcal K}\big)^{-1/2} \;,
\end{equation}
and finally
\begin{equation}\label{eq:gcdet}
\Gamma_\lambda({\mathcal C}) \;=\; \frac{1}{2}\,{\rm Tr} \log{\mathcal K} \;.
\end{equation}

We proceed in the next Section to perform a perturbative expansion of the
effective action in powers of the departure of the particle as measured
with respect to a static situation.

%%%%%%%%%%%%%%%%%%%%%%%%%%%%%%%%%%%%%%%%%%%%%%%%%%%%%%%%%%%%%%%%%%%%%
%%%%%%%%%%%%%%%%%%%%%%%%%%%%%%%%%%%%%%%%%%%%%%%%%%%%%%%%%%%%%%%%%%%%%
%%%%%%%%%%%%%%%%%%%%% Perturbative expansion %%%%%%%%%%%%%%%%%%%%%%%%
%%%%%%%%%%%%%%%%%%%%%%%%%%%%%%%%%%%%%%%%%%%%%%%%%%%%%%%%%%%%%%%%%%%%%
%%%%%%%%%%%%%%%%%%%%%%%%%%%%%%%%%%%%%%%%%%%%%%%%%%%%%%%%%%%%%%%%%%%%%

\section{Small-departure expansion for the massless field}\label{sec:pert}
We consider a worldline ${\mathcal C}$ parametrized with the `Lab' time
$\tau \equiv t \equiv x_0$. 
Therefore, $(y_\mu) = (y_\mu(t)) = (t, \eta_i(t))$ ($i=1,\ldots,d$),
and we assume that $\eta_i(t)$, the departure from a static situation:
$(t,{\mathbf 0})$, is small. By an adequate choice of the spatial origin,
we can always assume that the average position of the particle is ${\mathbf
0}$; thus: $\int_t \eta_i(t) =  0$.
Besides, we deal with non-relativistic motions, so that the
$\sqrt{g}$ factors will be replaced by $1$.

We proceed to perform an expansion in powers of $\eta_i(t)$.
Using an index to denote, in a given object, the order in $\mathbf
\eta_i$ in that expansion, we shall have:
\begin{equation}
	\Gamma_\lambda({\mathcal C}) \;=\; \Gamma^{(0)}_\lambda({\mathcal C}) 
	\,+\, \Gamma^{(1)}_\lambda({\mathcal C}) 
	\,+\,\Gamma^{(2)}_\lambda({\mathcal C})  \,+\, \ldots 
\end{equation}
where the explicit form of the first few terms is: 
\begin{align}
\Gamma^{(0)}_\lambda({\mathcal C})  &=\; \frac{1}{2}\,{\rm Tr} \log{\mathcal K}^{(0)}
\nonumber\\
\Gamma^{(1)}_\lambda({\mathcal C})  &=\; \frac{1}{2}\,{\rm Tr}\Big[ \big({\mathcal
K}^{(0)}\big)^{-1} \, {\mathcal K}^{(1)}\Big] \nonumber\\
\Gamma^{(2)}_\lambda({\mathcal C})  &=\; \frac{1}{2}\,{\rm Tr}\Big[ \big({\mathcal
K}^{(0)}\big)^{-1} \, {\mathcal K}^{(2)}\Big] 
  \,-\, \frac{1}{4}\,{\rm Tr}\Big[ \big({\mathcal
K}^{(0)}\big)^{-1} \, {\mathcal K}^{(1)}
\big({\mathcal K}^{(0)}\big)^{-1} \, {\mathcal K}^{(1)}\Big] \;.
\end{align}

It goes without saying that the $\Gamma^{(0)}_\lambda({\mathcal C})$,
independent of the particle's motion, may be safely discarded, and we shall
do so (it only contributes a constant to the static vacuum energy).

To evaluate the remaining terms, we need to consider the kernels ${\mathcal
K}^{(i)}$, $i=0,1,2$. 
We see that:
\begin{equation}
	{\mathcal K}_\lambda^{(0)}(t,t') \;=\; \frac{1}{\lambda} \, \delta(t-t') \,
+\, \langle t,{\mathbf 0} | (-\partial^2)^{-1} |
t', {\mathbf 0} \rangle \;=\; 
\int \frac{d\omega}{2\pi} \, e^{i \omega (t-t')} \, \widetilde{\mathcal
	K}_\lambda^{(0)}(\omega)  
\end{equation}
where:
\begin{equation}
	\widetilde{\mathcal K}_\lambda^{(0)}(\omega) \;=\; \frac{1}{\lambda} \,+\,	
I(\omega) \;,
\end{equation}
where
\begin{equation}
I(\omega) \;=\;\int \frac{d^dk}{(2\pi)^d} \frac{1}{{\mathbf k}^2 + \omega^2} 
\end{equation}
(we have used $\omega$ to denote the $k_0$ component of the momentum).

It is rather straightforward to see that ${\mathcal K}_\lambda^{(1)}$ vanishes
\begin{equation}
	{\mathcal K}_\lambda^{(1)}(t,t') \;=\; i \, \int \frac{d\omega}{2\pi} \,
e^{i \omega (t-t')} \int \frac{d^dk}{(2\pi)^d} 
\, \frac{1}{ {\mathbf k}^2+\omega^2} 
	\, k_j  (\eta_j(t)- \eta_j(t')) \;=\; 0 \;.
\end{equation}

Regarding ${\mathcal K}_\lambda^{(2)}$, we obtain: 
\begin{equation}
	{\mathcal K}_\lambda^{(2)}(t,t') \;=\; - \frac{1}{2} \,
\int \frac{d\omega}{2\pi} \, e^{i \omega (t-t')}
\int \frac{d^dk}{(2\pi)^d} 
\,\frac{k_i k_j}{ {\mathbf k}^2+\omega^2} 
\, (\eta_i(t)- \eta_i(t')) (\eta_j(t)- \eta_j(t')) \;.
\end{equation}
Or,
\begin{equation}
	{\mathcal K}_\lambda^{(2)}(t,t') \;=\; - \frac{1}{2 d} \,
\int \frac{d\omega}{2\pi} \, e^{i \omega (t-t')}
\int \frac{d^dk}{(2\pi)^d} 
\,\frac{{\mathbf k}^2}{{\mathbf k}^2+\omega^2} 
\, 
(\eta_i(t)- \eta_i(t'))^2 \;.
\end{equation}
We see that:
\begin{align}
	{\mathcal K}_\lambda^{(2)}(t,t') &=\;  \frac{1}{2 d} \,
\int \frac{d\omega}{2\pi} \, e^{i \omega (t-t')} \,\omega^2\,
I(\omega)  \, (\eta_i(t)- \eta_i(t'))^2 \nonumber\\
 &=\;\frac{1}{2 d} \,
	\big[ (\eta_i(t))^2 \,+\,(\eta_i(t'))^2 \big] \,
\int \frac{d\omega}{2\pi} \, e^{i \omega (t-t')} \,\omega^2\,
I(\omega) \nonumber\\ 
 &-\; \frac{1}{d} \, \eta_i(t) \,\eta_i(t') \,
\int \frac{d\omega}{2\pi} \, e^{i \omega (t-t')} \,\omega^2\,
I(\omega) \;. 
\end{align}

It is evident that the first-order term $\Gamma_\lambda^{(1)}$ vanishes.
Let us the calculate $\Gamma_\lambda^{(2)}$, the only surviving
contribution:
\begin{equation}
	\Gamma_\lambda^{(2)} \,=\, \frac{1}{2}\, {\rm Tr}( \Delta {\mathcal
	K}_\lambda^{(2)} )
\end{equation}	
with $\Delta \equiv ({\mathcal K}_\lambda^{(0)})^{-1}$.
Then
\begin{align}\label{eq:res1}
\Gamma_\lambda^{(2)} &=\, \frac{1}{2}\int_{t,t'}
	\Delta(t,t') {\mathcal K}_\lambda^{(2)}(t',t) \,=\, \int \frac{d\omega}{2\pi} \,[\widetilde{\mathcal
	K}_\lambda^{(0)}(\omega)]^{-1} e^{i \omega (t-t')} \, {\mathcal
	K}_\lambda^{(2)}(t',t)
	\nonumber\\
 &=\;\frac{1}{d} \,  \int_t (\eta_i(t))^2  \,
\int \frac{d\omega}{2\pi} \,[\widetilde{\mathcal
	K}_\lambda^{(0)}(\omega)]^{-1}  \omega^2\,
I(\omega) 
	\,+\,\frac{1}{2} \, \int \frac{d\omega}{2\pi}
	\,f(\omega) \,|\tilde{\eta}_i(\omega)|^2 
\end{align}  
where $\tilde{\eta}_i$ is the Fourier transform of $\eta_i$, and: 
\begin{equation}\label{eq:deff}
	f(\omega)\;=\; -\frac{1}{d} \int \frac{d\nu}{2\pi} \,
	[\widetilde{\mathcal K}_\lambda^{(0)}(\nu+\omega)]^{-1}  \nu^2\,
I(\nu) \;. 
\end{equation}
Note that the first term in the second line of (\ref{eq:res1}) is a mass
renormalization; we shall focus in what follows on the properties of the
second one. 
The treatment of such a term differs depending on the number $d$ of spatial
dimensions. Indeed, we see that $f$ depends on $I(\omega)$, both explicitly
and also through $\widetilde{\mathcal K}_\lambda^{(0)}$, and $I(\omega)$
diverges, except for $d=1$. We note that the very same divergence appears
when considering the $\delta$-function potential in $d > 1$. This requires
to renormalize the coupling $\lambda$, something which we will implement
here as well.

One can also see that, since ${\mathcal K}^{(1)}$ vanishes, the expression
for the fourth-order term simplifies to:
\begin{equation}
\Gamma^{(4)}_\lambda({\mathcal C})  \;=\; \frac{1}{2}\,{\rm Tr}\Big[ \big({\mathcal
K}^{(0)}\big)^{-1} \, {\mathcal K}^{(4)}\Big] 
  \,-\, \frac{1}{4}\,{\rm Tr}\Big[ \big({\mathcal
K}^{(0)}\big)^{-1} \, {\mathcal K}^{(2)}
 \big({\mathcal K}^{(0)}\big)^{-1} \, {\mathcal K}^{(2)}\Big] \;.
\end{equation}

\subsection{$d=1$}
The $d=1$ case has been previously studied~\cite{Fosco:2007nz}. In this
case, no renormalization of $\lambda$ is required, since the integral
$I(\omega)$ is convergent. Indeed,
\begin{equation}
	[ I(\omega) ]\large|_{d=1} \;=\; \frac{1}{2|\omega|} \;.
\end{equation}  
The $\nu$ integral in the expression for $f$ can then be explicitly
evaluated, the result being:
\begin{equation}
f(\omega) \,=\, - \frac{\lambda^2}{8\pi^2} \Big[
2 |\omega| - \lambda \big( 1 + \frac{2}{\lambda} |\omega| \big)
\ln\big(1 + \frac{2}{\lambda} |\omega|\big)\Big]\;.
\end{equation}
A large-$\lambda$ expansion of the previous expression yields
\begin{equation}\label{eq:flexp}
f(\omega) \,=\, \frac{\lambda}{4\pi} \omega^2
\,-\, \frac{1}{6\pi}  |\omega|^3 \,+\,{\mathcal O}(\lambda^{-1})\;,
\end{equation}
where one observes the different nature of the terms; the second one is the
well-known Dirichlet result, and the first one amounts to a renormalization
of the kinetic energy of the particle.

\subsection{$d=3$}
In $3$ spatial dimensions, we see that:
\begin{equation}
	[ I(\omega) ]_{d=3} \;=\; \frac{\Lambda}{2\pi^2} \,-\,
	\frac{|\omega|}{4\pi} \;.
\end{equation}
Inserting this into $\widetilde{\mathcal K}_\lambda^{(0)}$,
we now obtain instead
\begin{equation}
\widetilde{\mathcal K}_\lambda^{(0)}(\omega) \;=\; \frac{1}{\lambda_r} 
	 \,-\, \frac{|\omega|}{4\pi} \;,
\end{equation}
where: 
\begin{equation}
\frac{1}{\lambda_r} \;=\; \frac{1}{\lambda} \,+\,\frac{\Lambda}{2\pi^2} \;.
\end{equation}

Let us evaluate the kernel $f(\omega)$ for $d=3$, for the case $\lambda_r
\to \infty$. We see that:
\begin{equation}\label{eq:f31}
	f_D(\omega) \,\equiv\,	[f(\omega)]_{\lambda_r \to \infty}\;=\; -\frac{1}{3} \int \frac{d\nu}{2\pi} \,
	\frac{|\nu|^3}{|\nu + \omega|}\;. 
\end{equation}
The last integral may be obtained as:
\begin{align}
	f_D(\omega) &=\; 
	f(\omega;-3/2,1/2)  \nonumber\\
	f(\omega;\alpha_1,\alpha_2)] &=\; -\frac{1}{3} \int \frac{d\nu}{2\pi} \,
	\frac{1}{|\nu^2|^{\alpha_1}[(\nu + \omega)^2]^{\alpha_2}}\;,
\end{align}	
where, after a standard calculation, we find:
\begin{equation}
f(\omega;\alpha_1,\alpha_2)] \;=\; - 
\, 
\frac{\Gamma(\alpha_1 + \alpha_2 -1/2) [\Gamma(3/2 -\alpha_1 -
\alpha_2)]^2}{3 (4 \pi)^{1/2} \Gamma(\alpha_1) \Gamma(\alpha_2) \Gamma(3 -
2\alpha_1 - 2 \alpha_2)} \, |\omega|^{1 - 2 (\alpha_1+\alpha_2)} \;.
\end{equation}

Thus:
\begin{equation}
	f_D(\omega) \;=\;- \frac{1}{256} \, |\omega|^3 \;.
\end{equation}
We can also calculate explicitly the subleading terms. We have found that
all the terms involving even powers of the frequency $\omega$, and
therefore not contributing to the dissipative effects (imaginary part of
the analytically continued $\Gamma$), are divergent. The terms which are
odd in $|\omega|$, including the $\lambda_r \to \infty$ one, are finite:
\begin{equation}
	f(\omega) \;=\;-\frac{1}{256} \, |\omega|^3 \,+\, \frac{\pi^2}{\lambda_r^2} 
	\, |\omega|  \,-\, \frac{64 \pi^4}{ 3 \lambda_r^4} \,
	|\omega|^{-1} \,+\, {\mathcal O}(\frac{1}{\lambda_r^6}) \;. 
\end{equation}
Therefore, performing the rotation back to real time,
\begin{equation}
	{\rm Im}[ \Gamma_\lambda^{(2)}] \;=\;\frac{1}{2} \,
	\int\frac{d\omega}{2\pi} \, |\tilde{\eta}_i(\omega)|^2  \big[
		\frac{1}{256} \, |\omega|^3 \,-\, \frac{\pi^2}{\lambda_r^2} 
	\, |\omega|  \,+\, \frac{64 \pi^4}{ 3 \lambda_r^4} \,
	|\omega|^{-1} \;-\, \ldots \big] \;.	
\end{equation}

\section{The $d=2$ effective action and $s$-space renormalization}\label{sec:par}
\subsection{$m=0$}
In $d=2$,  we note that the integral in $I(\omega)$ is logarithmically
divergent. Introducing an UV cutoff $\Lambda$, we see that,
for large values of $\Lambda$,
\begin{equation}
	[ I(\omega)]_{d=2} \;=\; \frac{1}{2\pi} \,
	\log|\frac{\Lambda}{\omega}| \;.
\end{equation}
Following the usual treatment of the $\delta$-function potential, we introduce a
renormalization scale $\mu$, and rewrite
\begin{equation}\label{eq:iw2}
	[ I(\omega) ]_{d=2} \;=\; \frac{1}{2\pi} \,
	\log|\frac{\Lambda}{\mu}| \,-\, \frac{1}{2\pi} \,
	\log|\frac{\omega}{\mu}| \;.
\end{equation}
Inserting this into $\widetilde{\mathcal K}_\lambda^{(0)}$,
we see that
\begin{equation}
\widetilde{\mathcal K}_\lambda^{(0)}(\omega) \;=\; \frac{1}{\lambda_r} 
	 \,+\, I_\mu(\omega) \;,
\end{equation}
where we have introduced the renormalized coupling constant:
\begin{equation}
	\frac{1}{\lambda_r} \;=\; \frac{1}{\lambda} \,+\, \frac{1}{2\pi} \, \log|\frac{\Lambda}{\mu}|
	\;,
\end{equation}
and the scale-dependent function
\begin{equation}
I_\mu(\omega) \;=\;-\, \frac{1}{2\pi} \, \log|\frac{\omega}{\mu}| \;.
\end{equation}

Note that $I(\omega)$ also appears in the denominator of (\ref{eq:deff});
using (\ref{eq:iw2}), one sees that:
\begin{align}\label{eq:f21}
	f(\omega) &=\; - \frac{1}{4\pi} \log|\frac{\Lambda}{\mu}| \left[
	\, \int \frac{d\nu}{2\pi} \, 
	[\widetilde{\mathcal K}_\lambda^{(0)}(\nu)]^{-1} \nu^2\,+\, \omega^2  
	\, \int \frac{d\nu}{2\pi} \,
	[\widetilde{\mathcal K}_\lambda^{(0)}(\nu)]^{-1} \right]
\nonumber\\
	&+\, f_\mu(\omega) \;,
\end{align}
were the first two terms amount to a mass and kinetic term
renormalizations, while
\begin{equation}\label{eq:f22}
f_\mu(\omega)\;=\; -\frac{1}{2} \int \frac{d\nu}{2\pi} \,
[ \frac{1}{\lambda_r} + I_\mu(\nu+\omega)]^{-1}  \nu^2\,
I_\mu(\nu) \;, 
\end{equation}
which does not involve divergent objects in its integrand.

Let us now evaluate the previous integral, which may be rendered in the
following fashion:
\begin{equation}\label{eq:f23}
f_\mu(\omega)\;=\; -\frac{1}{2} \int_{-\infty}^{+\infty} \frac{d\nu}{2\pi} \,
	\frac{\log|\frac{\nu}{\mu}|}{\log|\frac{\nu+\omega}{\mu}
	e^{-\frac{2\pi}{\lambda_r}}|} \nu^2\, \;,
\end{equation}
where we have indicated the range of integration explicitly.

\begin{figure}[h]  
%\centering 
\subfloat[]{\includegraphics[scale=.59]{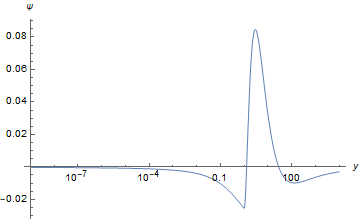}	}  \hspace{.0cm} 
	  	 \subfloat[]{\includegraphics[scale=.56]{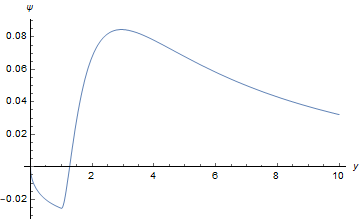} }                	  	               
\caption{ (a): $\psi(y)$ as a function of  $y \in [10^{-9}, 10^4]$; here we considered a logarithmic (linear) scale for the horizontal (vertical) axis. (b): $\psi(y)$ as a function of $y \in [10^{-9},
10]$ (with linear scales).}      
 \label{fig1}  
\end{figure}

\begin{figure}[h]  
%\centering 
      \subfloat[]{\includegraphics[scale=.5]{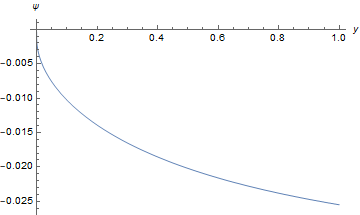} }   \hspace{.2cm}
      	  	 \subfloat[]{\includegraphics[scale=0.5]{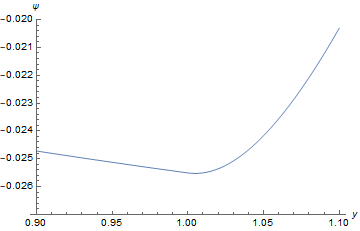} }  
\caption{ (a): A zoom of $\psi(y)$ in the interval  $y \in [10^{-9}, 1]$. (b): A zoom of $\psi(y)$ at the first local minimum of $\psi(y)$, which shows a regular slope.}      
 \label{fig2}  
\end{figure}

To perform the integral, we first perform a shift in the integration variable, and 
symmetrize it explicitly with respect to $\nu$. Then we obtain the equivalent expression:
\begin{equation}\label{eq:f24}
f_\mu(\omega)\;=\; -\frac{1}{4\pi} \int_0^{\infty} \,d\nu\,
	\left[\frac{\log|\frac{\nu + \omega}{\mu}|}{\log|\frac{\nu}{\mu} e^{-\frac{2\pi}{\lambda_r}}|} (\nu + \omega)^2 +\frac{\log|\frac{\nu - \omega}{\mu}|}{\log|\frac{\nu}{\mu} e^{-\frac{2\pi}{\lambda_r}}|} (\nu - \omega)^2 
	\right]\;. 
\end{equation}
The last integral is UV divergent; to cope with those divergences, we
subtract from the integrand its Taylor expansion around $\omega=0$, up to
the second order. This procedure does not erase information related to
dissipation, as the subtracted terms give rise to a renormalization of the
kinetic term and the mass of the particle. This leads (after some
algebra) to the subtracted integral $f_s$:
\begin{align}\label{eq:f25}
f_s(\omega) \;=\; -\frac{1}{4\pi} &\int_0^{\infty} \, \frac{d\nu}{\log|\frac{\nu}{\mu} e^{-\frac{2\pi}{\lambda_r}}|}
\Big[ \log\big| 1 - (\frac{\omega}{\nu})^2 \big| (\nu^2 + \omega^2) \nonumber\\
\,  & +\, 2 \, \log\big|\frac{\nu + \omega}{\nu - \omega}\big| \, \nu \, \omega  - 3 \, \omega^2 \Big]\;. 
\end{align}
The previous integral is UV-convergent. Performing a rescaling in the integration variable,
\begin{equation}\label{eq:f26}
f_s(\omega)\;=\; |\omega|^3 \; \psi\big(|\frac{\omega}{\mu}| e^{-\frac{2\pi}{\lambda_r}}\big) \;,
\end{equation}
where
\begin{align}\label{eq:f27}
\psi(y) \;=\; -\frac{1}{4\pi} &\int_0^{\infty} \, \frac{dx}{\log| x \, y |}
\Big[ (x^2 + 1) \, \log\big| 1 - \frac{1}{x^2} \big| \, \nonumber\\
\,  & +\, 2 \, x \, \log\big|\frac{x + 1}{x - 1}\big| \,  - 3  \Big]\;. 
\end{align}
$\psi(y)$ may be evaluated numerically, and it turns out to be finite and smooth for every $y>0$, as can be seen in Figs \ref{fig1} and \ref{fig2}, which were generated with \textsc{Mathematica}.

We have thus succeeded in renormalizing the codimension 2 case in $d=2$. 
Since $d=2$ exhibits interesting features, let us now consider, for the
sake of completeness, also a {\em massive\/} field with mass $m$.  

In what follows, we shall consider the
arc-length ($s$) of the curve, and introduce a regularization in
$s$-parameter space.
 Our starting point is Eq.  \eqref{eq:gcdet},
\begin{equation}
\Gamma_\lambda({\mathcal C})=\frac{1}{2} {\rm Tr}\,\log{(\hat{1} +\lambda{\mathcal H}_{\mathcal C})}+ {\rm const.}   \;,\label{effaction}
\end{equation}
where the kernel of the operator ${\mathcal H}_{\mathcal C}$, obtained from Eq. \eqref{Ker} as  
\begin{align}
{\mathcal H}_{\mathcal C}(s,s') =  \langle y(s) | (-\partial^2 + m^2)^{-1} |  y(s') \rangle =   \frac{1}{4\pi} \frac{e^{-m|y(s)-y(s')|}}{|y(s)-y(s')|}  \;,
\end{align}  
is singular when $s\rightarrow s'$. This object can be treated by means of a regularization which enables the identification of the singular part in the sense of distributions. For example, we could use dimensional regularization ($d=2-\epsilon$)  
and, of course, introduce a mass parameter $\lambda \to \lambda \mu^{\epsilon}$ so as to keep $\lambda $ adimensional:
\begin{align}
 {\mathcal H}_{\mathcal C}(s,s') \to  \frac{1}{(2\pi)^{\frac{3-\epsilon}{2}}}\left(\frac{m}{|y(s)-y(s')|}\right)^{\frac{1-\epsilon}{2}}K_{\frac{1-\epsilon}{2}}(m|y(s)-y(s')|) \;.
\end{align}
Another possibility is to consider the regularized quantity
\begin{align}
{\mathcal H}_{\mathcal C}(s,s') \to  {\mathcal H}^\epsilon_{\mathcal C}(s,s') = e^{-m|y(s)-y(s')|}  {\mathcal I}^\epsilon_{\mathcal C}(s,s')  \;,
\end{align} 
\begin{align}
 {\mathcal I}^\epsilon_{\mathcal C}(s,s') =     \frac{1}{4\pi} \frac{ \mu^{\epsilon}}{|y(s)-y(s')|^{1-\epsilon}} 
\label{Amass} \;.
\end{align}
In order to simplify the ($s$-independent) finite part, which will be absorved in the renormalized coupling constant, we shall adopt the second procedure. It is easy to see that in the limit $m \to 0$ this  coincides with the usual dimensional regularization.

\subsubsection{Regularization in $s$-space} 

We can initially rewrite
\begin{align}
{\mathcal H}^\epsilon_{\mathcal C}= D^\epsilon + {\mathcal H}^\epsilon_l \makebox[.5in]{,}   D^\epsilon(s,s') = {\mathcal H}^\epsilon_{\mathcal C} -
{\mathcal H}^\epsilon_l   
\label{Azero}
\end{align}
where ${\mathcal H}^\epsilon_l$ is the contribution of a line. In fact, the regulator can be removed in $D^\epsilon$, as $D = {\mathcal H}_{\mathcal C} -
{\mathcal H}_l $ is regular when $s\rightarrow s'$. To see this, we can expand $y(s')$ around $s$:
\begin{align}
y(s')-y(s) = y'(s)(s'-s)+\frac{y''(s)}{2}(s'-s)^2+\frac{y'''(s)}{3!}(s'-s)^3+\dots \;, \label{expr2}
\end{align}
and use that for the arc-length parameter it is verified ($e(s) = y'(s)$) 
\begin{equation}
|e(s)|^2=1 \makebox[.5in]{,} e(s)\cdot e'(s)=0 \makebox[.5in]{,}  e(s)\cdot e''(s)= - |e'(s)|^2 \;.
\end{equation}
 In other words,
\begin{align}
|y(s)-y(s')|=|s-s'|(1+h(s,s')) \makebox[.5in]{,} h(s,s') =- \frac{\dot{e}^2(s) }{24} \, (s-s')^2 + \dots 
\label{modex} 
\end{align} 
where the dots represent orders higher than $(s-s')^2$. Therefore, 
\begin{align}
D(s,s')=  \frac{1}{4\pi}\left(\frac{e^{-m|y(s)-y(s')|}}{|y(s)-y(s')|}-\frac{e^{-m|s-s'|}}{|s-s'|}\right) \;, \label{regular}
\end{align} 
is manifestly regular when $s$ approaches $s'$. In particular
\begin{align}
D(s,s)=\lim_{s\to s'} D(s, s') =0 \;.
\label{lim}
\end{align}
 Now, let us analyze 
\begin{align}
 {\mathcal H}_l^\epsilon (s,s') =  e^{-m|s-s'|}  {\mathcal I}^\epsilon_{l}(s,s')  \makebox[.5in]{,} 
   {\mathcal I}^\epsilon_{l}(s,s') =   \frac{1}{4\pi}\frac{\mu^\epsilon}{|s-s'|^{1-\epsilon}}  
   \label{Iele} \;,
%\label{Azero}
\end{align}  
which is the product of a regular factor times a distribution with singularities. It is well-known that the distribution $|x|^{\alpha}$ has a simple pole at $\alpha = -1$ whose residue is 
$2\delta(x)$ \cite{Gelfand}. Then, defining the dimensionless variable $x\equiv \mu (s-s')$, subtracting and adding the polar part of 
$ {\mathcal I}^\epsilon_{l}(s,s')$, and then multiplying by the regular factor $e^{-m|s-s'|}$, we get
\begin{align} 
 {\mathcal H}_l^\epsilon (s,s')= R^\epsilon(s,s') + \frac{1}{2\pi\epsilon} \, \delta(s-s')  \makebox[.5in]{,} R^\epsilon(s,s') = {\mathcal H}_l^\epsilon (s,s') - \frac{1}{2\pi\epsilon} \, \delta(s-s')   \;.
\label{regularizedkernel}
\end{align} 
We can check that  $R^\epsilon(s,s')$ is regular when $\epsilon \to 0$ by acting on a  test function
\begin{align}
& \int_{-\infty}^{+\infty} ds'\, R^\epsilon(s,s') f(s') =\int_{|s-s'|\leq \frac{1}{\mu}}ds' \,  \frac{1}{4\pi}\frac{\mu^\epsilon}{|s-s'|^{1-\epsilon}} (e^{-m|s-s'|}f(s')-f(s))\nonumber\\&+\int_{|s-s'|\geq \frac{1}{\mu}} ds' \,  \frac{1}{4\pi}\frac{\mu^\epsilon}{|s-s'|^{1-\epsilon}}  e^{-m|s-s'|}f(s')  \label{defR} \;.
\end{align}
Indeed, this is well-defined in the limit $\epsilon \to 0$. Then, introducing the renormalized coupling constant $\lambda_r$,
\begin{align}
&\frac{1}{\lambda_r}=\frac{1}{\lambda}+\frac{1}{2\pi \epsilon} \;,
\label{lambdar}
\end{align}
 the contribution to the effective action is obtained from the $\epsilon \to 0$ limit of
\begin{align}
\Gamma_\lambda({\mathcal C})= \frac{1}{2}{\rm Tr}\log{\left(\frac{1}{\lambda_r}+D+R^\epsilon\right)} \;,
\label{regulareff}
\end{align}
up to an irrelevant constant.

\subsubsection{The weak $\lambda_r$-limit}

We may perform an expansion for small $\lambda_r$:
\begin{align}
\Gamma_\lambda({\mathcal C}) = \delta\Gamma^{(1)}_\lambda({\mathcal C})+\delta\Gamma^{(2)}_\lambda({\mathcal C})+\dots .
\end{align}
The first and second order contributions read, respectively,
\begin{eqnarray}
 \delta\Gamma^{(1)}_\lambda({\mathcal C}) &=& \frac{\lambda_r}{2}\int ds\, D(s,s)+\frac{\lambda_r}{2}\int ds \, R^\epsilon(s,s)  \;, \label{firstorder}  \\
\delta\Gamma^{(2)}_\lambda({\mathcal C})&=&  -\frac{\lambda_r^2}{4}\int ds ds' \, \left(D^2(s,s')+2D(s,s')R^\epsilon(s',s)+{R^\epsilon}^2 (s,s') \right) \;. 
 \label{secorder}  
\end{eqnarray}  
Because of Eq. \eqref{lim}, the first term in Eq. \eqref{firstorder} vanishes, while the second term, in spite of the regularization, is still an ill-defined divergent quantity.  The same happens with the $R^2$-contribution in Eq. \eqref{secorder}. However, any improved regularization that keeps the natural dependence of $R$ in $s-s'$, which represents the translation symmetry of the contribution along the line, will give a divergence proportional to the length of the curve $ \int ds = L \to \infty$. 
This is associated to a renormalization of the string tension.  In the $\epsilon \to 0$ limit, the cross term in Eq. \eqref{secorder}, integrated over $s'$, gives (cf. Eq. \eqref{defR})
\begin{align}
&-\frac{\lambda_r^2}{2}\int ds ds'\,R(s,s')D(s,s')=-\frac{\lambda_r^2}{2}\int_{-\infty}^\infty ds\int_{|s-s'|\geq \frac{1}{\mu}} ds' \, \frac{1}{4\pi}\frac{e^{-m|s-s'|}}{|s-s'|}  D(s',s) \nonumber\\&-\frac{\lambda_r^2}{2}\int_{-\infty}^\infty ds\int_{|s-s'|\leq \frac{1}{\mu}} ds' \,\frac{1}{4\pi}\frac{1}{|s-s'|} (e^{-m|s-s'|}D(s',s)-D(s,s))\;. 
\end{align}
This expression is, by construction, regular. Since $D (s,s)=0$, this can be written in the simpler form:
\begin{align}
-\frac{\lambda_r^2}{2}\int ds ds' \,\frac{1}{4\pi}\frac{e^{-m|s-s'|}D(s,s')}{|s-s'|} \;.
\end{align}
Note that $D(s,s')\geq 0$, so that this is a negative contribution to the effective action, as well as that originated from the $D^2$-term.

%%%%%%%%%%%%%%%%%%%%%%%%%%%%%%%%%%%%%%%%%%%%%%%%%%%%%%%%%%%%%%%%%%%%%
%%%%%%%%%%%\lambda_r\to \infty%%%%%%%%%%%%%%%%%%%%%%%%%%%%%%%%%%%%%%%%%%%%%%%%%%%
%%%%%%%%%%%%%%%%%%%%%%%%%%%%%%%%%%%%%%%%%%%%%%%%%%%%%%%%%%%%%%%%%%%%%
%%%%%%%%%%%%%%%%%%%%%%%%%%%%%%%%%%%%%%%%%%%%%%%%%%%%%%%%%%%%%%%%%%%%%

\subsubsection{The small curvature limit}
\begin{figure}[h]
\centering
\includegraphics[scale=1]{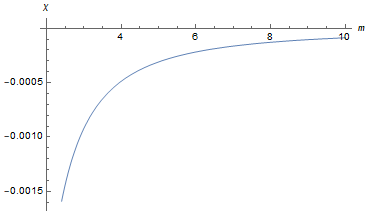}
\caption{$\chi$ as a function of $m$, for $\mu=1$ and large $\lambda_r$.}
\label{chi}
\end{figure}

An interesting physical situation to analyze is when the acceleration of the particle is small, so that, due to the mass gap, it does not radiate. To obtain the lowest order contribution of acceleration, it will be useful to perform an expansion of $\Gamma_\lambda(\mathcal{C})$ in powers of $D$, which tends to zero when ${\mathcal C} \to l$. For this objective, we can rewrite 
\begin{align}
& \Gamma_\lambda({\mathcal C})=\frac{1}{2}{\rm Tr}\log{\left(\lambda_r^{-1} +R^\epsilon\right)} + \frac{1}{2}{\rm Tr}\log{\left( 1+ \left(\lambda_r^{-1}+R^\epsilon\right)^{-1} D \right)} \nonumber \\
&  \Gamma_\lambda({\mathcal C})=\Gamma_\lambda(l)  + \frac{1}{2}{\rm Tr} {\left(   \left( \lambda_r^{-1} +R^\epsilon\right)^{-1} D \right)} -\frac{1}{4} {\rm Tr} {\left( \left( \lambda_r^{-1} +R^\epsilon\right)^{-1}D \,   \left(\lambda_r^{-1} +R^\epsilon\right)^{-1} D \right)}+\dots \;. \nonumber
\end{align}
Let us analyze the second term,
\begin{align}
 \frac{1}{2}{\rm Tr} {\left(   \left(\frac{1}{\lambda_r}+R^\epsilon\right)^{-1}D \right)} =  \frac{1}{2}\int ds \int ds' \, Q(s-s') \, D (s,s')
\end{align}
where $Q(s-s')$ is the kernel of the operator $ \left(\lambda_r^{-1}  +R^\epsilon\right)^{-1}$.  The term proportional to $\dot{e}^2(s)$ can be obtained by using Eq. \eqref{modex}
\begin{align}
& \frac{e^{-m|y(s)-y(s')|}}{|y(s)-y(s')|}  =  \frac{e^{-m|s-s'|}}{|s-s'|} \frac{e^{
\frac{m}{24} \dot{e}^2 |s-s'|^3 } }{(1 -
\frac{1}{24} \dot{e}^2 |s-s'|^2 )} + \dots \nonumber \\
&   =  \frac{e^{-m|s-s'|}}{|s-s'|}  + \dot{e}^2(s) \, P(s-s')+\dots \;, \nonumber \\
&  P(s-s') =  \frac{1}{24}   ( |s-s'| + m 
 |s-s'|^2 ) \, e^{-m|s-s'|} \makebox[.5in]{,}  D(s,s')=\frac{\dot{e}^2(s)}{4\pi}  \, P(s-s') \;,
\end{align} 
\begin{align}
 {\rm Tr} {\left( (\lambda_r ^{-1} + R^\epsilon)^{-1} D \right)} = \int ds \frac{\dot{e}^2(s)}{4\pi} \int ds' \, 
 Q(s-s') \, \, P(s-s')  \nonumber \\
  =  \int ds \frac{\dot{e}^2(s)}{4\pi} \int du \, Q(u)  P(u)   \;.
\end{align} 
Now, if the kernel of the operator $R^\epsilon$ is $R^\epsilon(s-s')$, then, in terms of the Fourier transforms
\begin{align}
P(u) = \int \frac{d\omega}{2\pi} \,  \tilde{P}(\zeta) e^{i \zeta u}  \makebox[.5in]{,} R^\epsilon(u) = \int \frac{d\zeta}{2\pi} \,  \tilde{R}^\epsilon(\zeta) e^{i \zeta u} 
\end{align}
we have
 \begin{align}  
\int du \, Q(u)  P(u) = \int \frac{d\zeta}{2\pi}\,  \frac{\tilde{P}(\zeta)}{\frac{1}{\lambda_r}+\tilde{R}^\epsilon(\zeta)} \;.
\label{essentially}
\end{align}
$P(\omega)$ is  found to be
\begin{align}
\int_{-\infty}^{+\infty}du\,\frac{1}{24}\left(|u|+m|u|^2\right)e^{-m|u|}e^{-i\zeta u}=\frac{3m^4-6m^2\zeta^2-\zeta^4}{12(m^2+\zeta^2)^3} \;.
\end{align}
The explicit form of $R^\epsilon(s-s')$ is obtained from Eqs. \eqref{Iele}, \eqref{regularizedkernel}
 \begin{align}
R^\epsilon(s-s') = \frac{\mu^\epsilon}{4\pi}\frac{e^{-m|s-s'|}}{|s-s'|^{1-\epsilon}} - \frac{1}{2\pi \epsilon}  \delta(s-s')  \;.
\end{align}
Performing the Fourier transform, using that for $\epsilon >0$ \cite{Gelfand}  
\begin{align}
{\mathcal F}(e^{-m|x|}|x|^{\epsilon-1})={\mathcal F}(e^{-mx}x_+^{\epsilon-1})+{\mathcal F}(e^{mx}x_-^{\epsilon-1})= ie^{i(\epsilon-1)\frac{\pi}{2}}\Gamma(\epsilon)(-\zeta+i m)^{-\epsilon}+ {\rm c.c.} \;, \label{transform}
\end{align}
we arrive at
\begin{align}
\tilde{R}^\epsilon(\zeta)={\mathcal F}(R^\epsilon)(\zeta)=-\frac{\gamma}{2\pi}-\frac{1}{4\pi}\log{\frac{(\zeta^2+m^2)}{\mu^2}}+O(\epsilon) \;.
\end{align}
Here, we can safely take the limit $\epsilon \to 0$ to conclude  
\begin{align}
\lim_{\epsilon\to 0^+}\tilde{R}^\epsilon(\zeta)=-\frac{\gamma}{2\pi}-\frac{1}{4\pi}\log{\frac{(\zeta^2+m^2)}{\mu^2}} \;.
\end{align}
Finally, to get rid of $\gamma$, we redefine $\mu\rightarrow \mu e^{-\gamma}$, 
which implies 
 \begin{align}  
\int du \, Q(u)  P(u) =\frac{1}{6}\int_{-\infty}^{+\infty} d\zeta\,  \frac{6m^2\zeta^2+\zeta^4-3m^4}{(m^2+\zeta^2)^3\left(-\frac{4\pi}{\lambda_r}+\log{\frac{\zeta^2+m^2}{\mu^2}}\right) } \;.
\end{align}
Then, to lowest order, we find 
\begin{align}
&\Gamma_\lambda({\mathcal C})-\Gamma_\lambda(l) =\chi(m,\mu) \int ds\, \dot{e}^2(s) ,\\
&\chi(m,\mu)=\frac{1}{96\pi}\int_{-\infty}^{+\infty} d\zeta\,  \frac{6m^2\zeta^2+\zeta^4-3m^4}{(m^2+\zeta^2)^3\log{\big( \frac{\sqrt{\zeta^2+m^2}}{\mu} e^{-\frac{2\pi}{\lambda_r}}}\big)},
\end{align}
where $\Gamma_\lambda(l)$ is the effective action of a straight line, and
$\chi (m,\mu)$ is a constant that depends on the mass and the arbitrary
scale $\mu$. 
For $m >> \mu\, e^{\frac{2\pi}{\lambda_r}}$, $\chi$ is finite and negative. In Fig. \ref{chi}, we show the dependence of the $\chi$-coefficient with $m$ in this regime.

\section{Conclusions}\label{sec:conc}
In this work we have defined and studied quantum dissipation in a moving 
DCE setting involving $\delta$ interactions with codimension larger than one.  
Specifically, we studied a real scalar field in $d+1$ dimensions ($d=2,3$)
coupled to an imperfect one dimensional mirror. We found that the singular
nature of the problem requires a renormalization of the coupling
($\lambda$) between the field and the mirror. In particular, for $d=3$,
there is a finite scale-independent imaginary part for the effective
action. In the Dirichlet limit, similarly to the well-known $d=1$ result,
this imaginary part
contains a $|\omega|^3$ dependence, which is what one expects on
dimensional grounds, assuming no renormalization scale dependence is
generated.

The case $d=2$, where $\lambda$ is dimensionless, is special, for the
coupling not only gets renormalized but also acquires a dependence on a
mass scale $\mu$. This phenomenon parallels that observed when dealing with
the $\delta$-potential in quantum mechanics for a planar system
\cite{Jackiw:1991je}. In this case, we have shown how the mass scale $\mu$ is
generated for the massless field, and how it intervenes in the construction
of the renormalized  effective action, in the small-departure approximation.  
We have also found that, for a massive field in $d=2$, apart from inducing 
renormalizations for the kinetic energy and the mass of the mirror, the Euclidean effective action
is finite and negative for small $\lambda_r$. Finally, we considered a
massive field coupled to an imperfect mirror with small acceleration, where
no imaginary part is expected, and found that the effective action is lower
than that of a static mirror.  

It would be very interesting to extend this analysis to other cases with
codimension 2 that arise when considering quantum fluctuations around
vortex-like defects in three and four Euclidean dimensions.  In particular,
it would be important to study the singular problem associated with curved
thin center vortices in Yang-Mills theories, 
and obtain quantum properties such as stiffness from a fundamental point of view.

%In the context of calculating the effective action with a one-dimensional thin center vortex as a background, this would imply in a negative stiffness, which is in accordance with the result for $2$ dimensional vortices in $3+1$ dimensions \cite{hugo} 

\section*{Acknowledgements}
The Conselho Nacional de Desenvolvimento Cient\'{\i}fico e Tecnol\'{o}gico
(CNPq), ANPCyT, CONICET and UNCuyo are acknowledged for  financial support.

\end{document}